\documentclass{kapproc} % Computer Modern font calls
\usepackage{graphicx}

\kluwerbib
\begin{document}

%------------ article title  ------------------->>

\articletitle[]{The Galaxy Cluster RBS380:\\X-ray and optical analysis}

%% optional, to supply a shorter version of the title for the running head:
\chaptitlerunninghead{RBS380 Galaxy Cluster}

\author{R. Gil-Merino}
\affil{Institut f\"{u}r Physik, Universit\"{a}t Potsdam\\
Am Neuen Palais 10, D-14469 Potsdam, Germany}
\email{rmerino@astro.physik.uni-potsdam.de}

\author{S. Schindler}
\affil{Institut f\"{u}r Astrophysik, Universit\"{a}t Innsbruck\\
Technikerstr. 25, A-6020 Innsbruck, Austria}
\email{Sabine.Schindler@uibk.ac.at}

\begin{abstract}
We present X-ray and optical observations of the z$=$0.52 galaxy cluster RBS380. This is
the most distant cluster in the ROSAT Bright Source catalog. The cluster was observed with the CHANDRA
satellite in September 2000. The optical observations were carried out with the NTT-SUSI2
camara in filters V and R in August and September 2001. The
preliminary conclusions are that we see 
a very rich optical galaxy cluster but with a relative low X-ray luminosity. We also compare
our results to other clusters with similar properties.
\end{abstract}

%------------ body of article ------------------->>
\section{Introduction}
The galaxy cluster RBS380 was observed with NTT-SUSI2 as part of a
programme to search for gravitational arcs in X-ray bright clusters of
galaxies. We selected the most X-ray luminous clusters from the ROSAT
Bright Survey (Schwope et al. 2000), which have a predicted probabilty
for arcs of 60\%. RBS380 is the most distant
cluster of this sample and is therefore the first one that we
analyse. In a parallel programme several of these clusters are
observed in X-rays with CHANDRA. The aims of the project are
threefold. One aim is to compare
mass determinations by the gravitational lensing method and the X-ray
method. A second aim is to perform arc statistics to constrain
cosmological models. A third aim is to study in detail the distant background
galaxies that are magnified by the gravitational lensing effect.

\section{Observations and Data Reduction}

\subsection{Optical Data}
The galaxy cluster RBS380 was observed with the New Technology Telescope (NTT) 
in service mode during the Summer 2001. The Superb Seeing Imager-2 (SUSI2)
camera was used in bands V and R. The SUSI2 detector is a 2 CCDs array, 
each with 1024$\times$1024 pixels, subtending an area on the sky
of $5'.5\times5'.5$.\\
The data reduction was performed with the IRAF package. A total number of 6
images with very good seeing conditions ($\leq1''$) were used in the analysis, with an
exposure time of ~760 sec each.
After bias subtracting, the flatfielding cannot be done in a
standard way. A hyper-flat (Hainaut et al. 1998) was built using both the
twilights flats set and the night-sky (scientific) images set. The twilights
correct very well the high spatial frecuencies in the field, whereas from the
scientific images one obtains the low spatial frecuencies correction. We found
that a simple linear combination of the two final flats (one from the twilight
flats and the other from the scientific images) was enough for producing a high
quality final one. Of course, the procedure has to be done for each filter.\\
Once the images were flatfielded, they were co-added, obtaining a deep image of
the field and free of gaps between the chips (this was possible due to the
dithering between the exposures).

\subsection{X-ray Data}
The RBS380 cluster was observed in October 17, 2000 by the CHANDRA X-ray
Observatory (CXO). A single exposure of 10.3 ksec (~3 hrs) was taken with the Advance CCD
Imaging Spectrometer (ACIS). Only the 2$\times$2 arrays ACIS-I chipset was active. Each
CCD is a 1024$\times$1024 pixels array. The pixel resolution is $~0''.5$, covering a
total area on the sky of $~17'\times17'$.\\
The data were processed with the CIAO suite toolkit provided by the CXC. We
upgraded the gainmaps, checked the correction for the aspects offsets and
removed bad pixels from the field (using the files provided by the CXC). Then we
built the lightcurve for the observing period, looking for short high
backgrounds intervals, but finding none.\\
Since we are interested in diffuse emission, the first step is removing the
point sources present. We used the \emph{wavdetect} subroutine to detect them.
We applied it to a broadband (0.3-10 keV) image. We found 29 sources, expecting
none to be spurious. After removing the sources, we filled the holes using the
same counts distribution of background pixels for that region.\\
The correction for telescope vignetting and variations in the spatial efficiency
of the CCDs was done by means of an exposure map. The exposure map was generated
for a integrated energy distribution peak of 0.7 keV. Background correction was
also applied by using the sets provided by the CXC.

\section{Results}
\begin{figure}[hbtp]
 \centering
 \includegraphics[bbllx=204,bblly=298,bburx=406,bbury=496,width=5.0cm,
                  angle=0,clip=true]{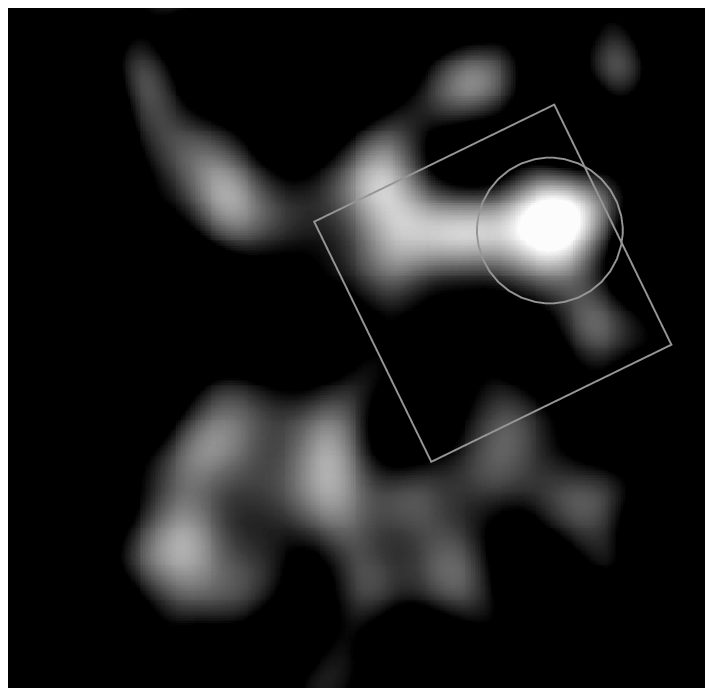}
 \includegraphics[bbllx=188,bblly=284,bburx=424,bbury=510,width=5.0cm,
                  angle=26,clip=true]{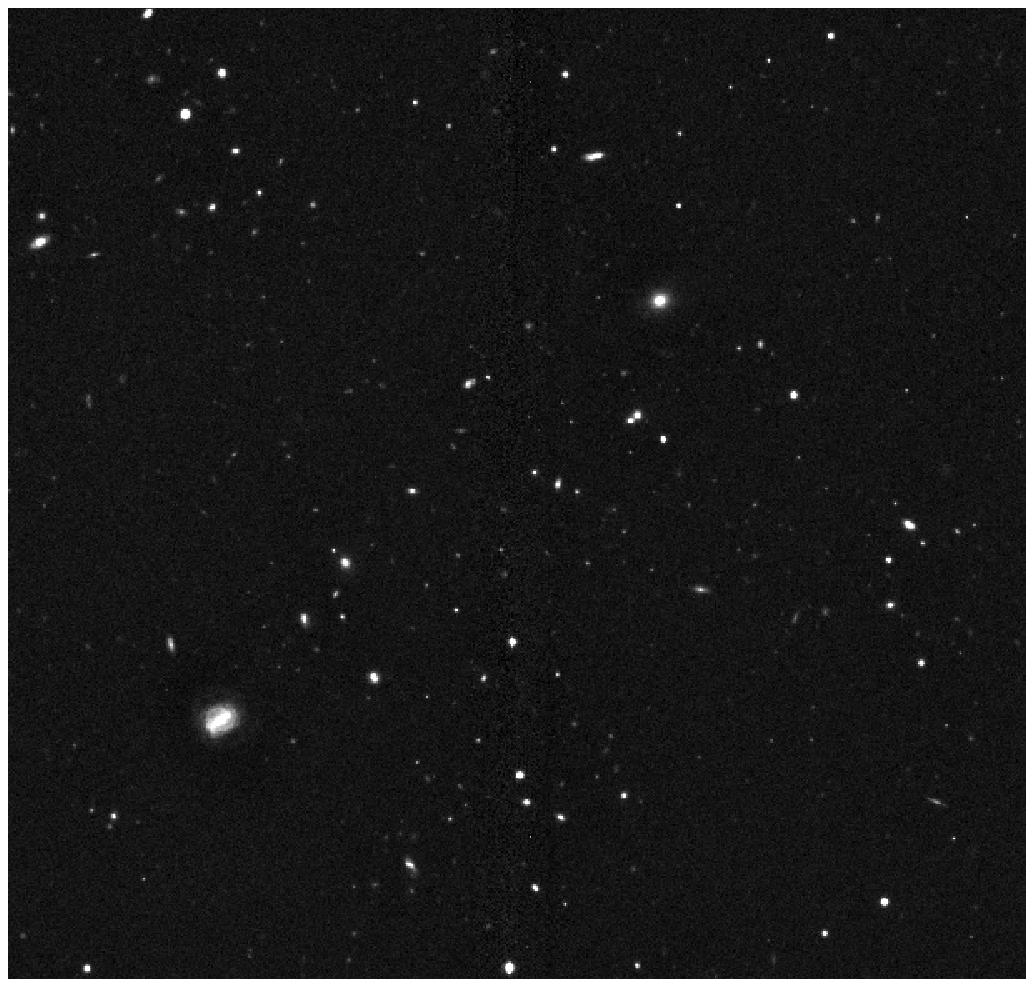}			
 \caption[]{\emph{Leftside}: X-ray image of the cluster. The total area is
 $14'\times14'$. The rotated square shows the region that was observed in the
 optical band. The circle has a radio of $1'.5$, the area where we computed the 
 count rate. \emph{Rightside}: The optical counterpart of the selected region.
 The total area is $~5'\times5'$. Although the number density of galaxies seems to
 be high, no gravitational arcs were detected. In both panels North is up and
 East is left.}
 \label{cumulo} 
\end{figure}

The preliminary results show optical R and V images with a high number density of
galaxies. The right side of Fig.~\ref{cumulo} shows the final R band
image. So far
no gravitational arcs have been detected. On the left side of Fig.~\ref{cumulo}
we present the results in X-ray. We selected a circular region of radius $1'.5$
centred at the peak of the emission. We obtain a count rate of 0.05 counts/s in the
encircled region, where most of the cluster emission is supposed to be. The derived
X-ray luminosity is $L_X=1.06~10^{44}$ erg/s. This is a relatively low X-ray luminosity
for a massive cluster of galaxies.\\
%%------------------------
%%--------------------------
Furthermore, the structure found in the X-ray emission is not perfectly symmetrical,
but presents extended emission to both sides of the main region. This could mean that
the cluster is not relaxed, but rather interacting with surrounding material or
another close cluster. This is point that merits more attention indeed.

\section{Conclusions}
A clear correlation between optical richness and X-ray luminosity is not yet well
established. Although it is in general assumed that a high number density of galaxies
should imply a great amount of intergalactic hot gas, there are few cases in the
literature that show little correlation between these two facts. In particular,
Cl0500$-$24 and Cl039$+$4713 are two interesting cases (Schindler \& 
Wambsganss 1996, 1997; Schindler et al. 1998). In Table~\ref{cuadro} we show the X-ray luminosities of the 
three clusters.\\
%%--------------
\begin{table}[tb]
\centering
\begin{tabular}{lllll}
\hline\noalign{\smallskip}
Name & Redshift & Luminosity [erg/s] & band\\
\noalign{\smallskip}\hline\noalign{\smallskip}
Cl0500$-$24   & 0.32 & 5.6~10$^{44}$  & bolometric \\
Cl0939$+$4713 & 0.41 & 7.9~10$^{44}$  & bolometric \\
RBS380        & 0.52 & 1.06~10$^{44}$ & [0.3-10] keV \\
\noalign{\smallskip}\hline
\end{tabular}
\caption{We compare the X-ray luminosity of RBS380 with two more clusters of galaxies
which are also optically rich, but have relatively low X-ray luminosity.}
\label{cuadro}
\end{table} 
%%--------------
Since we are presenting preliminary results, we have not fully analysed the optical
images of RBS380. In a forthcoming paper we will build a catalogue of objects in the
field. With such a catalogue we will be able to establish what are the galaxies that
belong to the cluster and get a precise number density map.

%------------ end of article ------------------->>

%% optional
%\section{Summary}

%% optional
\begin{acknowledgments}
This work has been partly supported by a predoctoral Marie Curie Fellowship to
RGM at the Liverpool John Moores Astrophysics Institute (HPMT-CT-2000-00136).
We specially thank Betty De Filippis and Africa Castillo for
many useful discussions and time-sharing. We also thank Olivier Hainaut for
detailed explanations on hyper-flatfielding.
\end{acknowledgments}

\end{document}